# Korean journals in the Science Citation Index:
# What do they reveal about the intellectual structure of S&T in Korea?

Han Woo PARK
YeungNam University, Korea (South)
hanpark@ynu.ac.kr; parkhanwoo@hotmail.com
http://www.hanpark.net
&
Loet Leydesdorff
Amsterdam School of Communications Research (ASCoR), University of Amsterdam
Kloveniersburgwal 48, 1012 CX Amsterdam, The Netherlands;
loet@leydesdorff.net; http://www.leydesdorff.net

**Abstract**

During the last decade, we have witnessed a sustained growth of South Korea's research output in terms of the world share of publications in the Science Citation Index database. However, Korea's citation performance is not yet as competitive as publication performance. In this study, the authors examine the intellectual structure of Korean S&T field based on social network analysis of journal-journal citation data using the ten Korean SCI journals as seed journals. The results reveal that Korean SCI journals function more like publication places, neither research channels nor information sources among national scientists. Thus, these journals may provide Korean scholars with access to international scientific communities by facilitating the respective entry barriers. However, there are no citation relations based on their Korean background. Furthermore, we intend to draw some policy implications which may be helpful to increase Korea's research potential.

**Keywords**: Korea, science citation index, national journals, social network analysis, science and technology, citation network

## Introduction

During the last decade, we have witnessed a sustained growth of South Korea's research output in terms of the world share of publications in the Thomson ISI database. According to the Korean Institute of Science & Technology Evaluation and Planning (KISTEP), there were 23,048 publications with at least one Korean address among the authors in the *National Science Indicator* 2006 of the Thomson ISI (KISTEP, 2006). Using the web-version of the *Science Citation Index*, we find higher numbers, but the linearly upward trend is very clear (Figure 1); Korea increases its percentage world share of publications with approximately 0.20 percent point each year ($r^2 > 0.99$). In terms of the number of papers in the *Science Citation Index* journals, Korea occupied the 14th position in the year 2005. This means a jump from 21st place in the year 1996. In other words, the *Science Citation* contained about four times as many publications with a Korean address in 2006 (28,059) as in 1996 (7,158). Figure 1 shows that South Korea has gone up in percentage world share of publications from 0.99% in 1996 to 2.86% in 2006.

**Figure 1.** Long-term trend of the percentage publications with a Korean address in the *Science Citation Index* (expanded version).

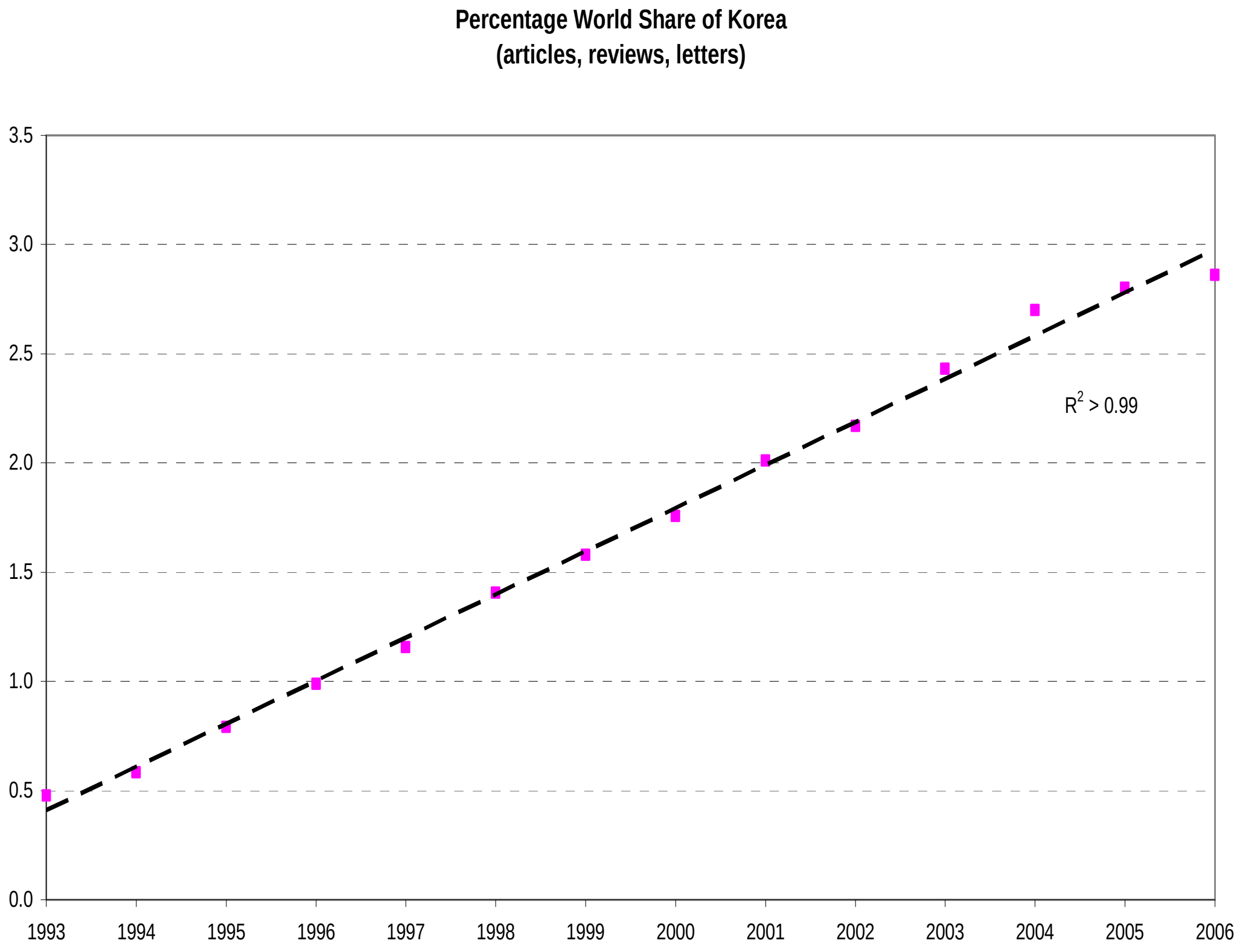

According to Kostoff (2004), Korea obtained even the 6th position during the first eight months of the year 2004 in the field of nanotechnology (Leydesdorff & Zhou, 2007). The KISTEP (2006) reports that molecular biology and genetics are by far the most influential areas of Korean science when measured by the citation index from the year 1996 to 2005. The next highly cited subjects are immunology, space science, neurology & behavioral science, biology & biochemistry. New fields including molecular biology and space engineering are significantly emerging in Korea.

Using a case study approach, Choung, Min, & Park (2003) showed that research capabilities in the information and telecommunication sector have improved considerably via domestic and international collaboration as well as governmental development polices. In another context, Leydesdorff and Zhou (2005) argued that Korea is one among five Asian nations that show a spectacular increase in their publication and citation rates when two periods (1997-2001 and 1993-1997) are compared at the level of individual nationals (King, 2004). Korea is from this perspective an interesting case. It is an OECD (Organization for Economic Cooperation and Development) member state and part of the western system. Nevertheless, the rise in publications etc. shows some “Chinese” patterns.

However, citation performance in terms of the percentage of world share, that is, the number of citations per paper, is not yet as competitive as publication performance. In a comparison between South Korea and the Netherlands based on the *Science Citation Index* 2002, Park, Hong, and Leydesdorff (2005) found that some traditional areas such as Korean chemistry are among the mainstream subjects in the increasing scientific and technological outputs. While research in the university sector has been well developed, industrial and public sector research efforts were relatively underrepresented in the database. The authors suggest that the participation of industry and public sectors in a university-driven knowledge production system can function as the crucial variable for boosting Korea’s research and citation performance.

These conflicting portfolios have motivated us to examine in more detail Korean scientific journals as one of the main output channels of this national research system. Since the knowledge production system is largely university-based, this question has also policy relevance in an emerging knowledge-based economy. Specifically, we analyze the detailed citation environment of international journals (operationalized as journals included in the *Science Citation Index*) published in Korea and compare them with international journals published elsewhere. Do Korean journals provide a specific communication channel both nationally and internationally? Do they add possibilities for Korean researchers or are Korean researchers publishing in these journals locking themselves into a national publication circuit? (Zhou & Leydesdorff, 2007). The results from this empirical exercise can be useful for academic actors (e.g., authors, journal editors and publishers, research policy agencies). Furthermore, we intend to draw some policy implications which may be helpful to increase Korea’s research potential.

**Korean journals in the national research system**

The majority of Korean scholarly journals are university-based. Korean journals have been mostly published solely by academic societies without cooperation from commercial publishers. Thus, there is always a minimum charge for the printing service. It has been hard for the national societies to raise sufficient funds for these publication activities. Authors with external funding are sometimes charged an additional fee because their research has more financial resources. In other words, Korea has a self-supporting system of publications. This is different from western countries where world-wide publishing companies are influential in the scientific publication market.

Table 1 summarizes the Korean journals included in the *Science Citation Index* that will be the subject of this study. The table also shows when they were first indexed in the Thomson ISI database. Among these journals, only *ETRI Journal* has a governmental affiliation. This journal is published by the Electronics and Telecommunications Research Institute (ETRI) that is sponsored by the Ministry of Information and Communication. Appendix I provides a similar list using the web-based version of the *Science Citation Index-Expanded.*

Table 1. A profile of the 10 Korean journals included in the *Science Citation Index* in 2004.

| No | Journal title (Abbreviation) | Publisher | Since when included in the *SCI* |
|---|---|---|---|
| 1 | *Bulletin of the Korean Chemical Society (B Kor Chem Soc)* | Korean Chemical Society | 1981 |
| 2 | *Journal of the Korean Physical Society (J Korean Phys Soc)* | Korean Physical Society | 1993 |
| 3 | *Molecules and Cells (Mol Cells)* | Korean Society for Molecular and Cellular Biology | 1995 |
| 4 | *ETRI Journal (Etri J)* | Electronics and Telecommunications Research Institute | 1996 |
| 5 | *Journal of Microbiology and Biotechnology (J Microbiol Biotechn)* | Korean Society for Microbiology and Biotechnology | 1995 |
| 6 | *Experimental and Molecular* | Korean Society of Medical | 1996 |
| | *Medicine (Exp Mol Med)* | Biochemistry and Molecular Biology | |
| 7 | *MacromolecularResearch (MacromolRes)(* Formerly *Korean Polymer, Korea Polym J*) | Polymer Society of Korea | 1995 |
| 8 | *Journal of Communications and Networks (J Commun Netw – S Kor)* | Korean Institute of Communication and Sciences | 2000 |
| 9 | *Journal of Korean Medical Science (J Korean Med Sci)* | Korean Academy of Medical Sciences | 1999 |
| 10 | *Journal of Ceramic Processing Research (J Ceram Process Res)* | Ceramic Processing Research Center of Hanyang University | 2002 |

- The abbreviations used for journal titles are available at http://apps.isiknowledge.com/WoS/help/A_abrvjt.html

Korean "*SCI*-journals" can be expected to have specific roles in the national research system. The internationalization of Korean research is promoted by these journals in several ways. Prospective authors are encouraged to submit their papers in English. The editors need to comply to the internationally-acceptable standards in reviewing procedures before the journal can be included in the Thomson ISI database. The quality of published research article needs to be controlled in order to maintain the status once the journal is indexed. Korean SCI journals also function as a nesting place where national scientists seek to publish preliminary research results that could be developed to be submitted into prestigious international journals.

Furthermore, national researchers consider Korean SCI journals as a PR (Public Relations) channel in order to get their own research exposed to peer-scientists, university administrators, and funding agencies. The scholarly discourse among elite Korean scientists is mediated through these journals. In addition, universities in Korea have become very strict in promoting faculty members these days. Several major schools require young scholars to publish their research output only in the SCI journals if they want to continue their employment contract and get promoted. Therefore, the Korean journals listed in the SCI provide national scientists with a nurturing environment both academically and administratively.

Kim and Kim (2000) did a bibliometric analysis of publications by chemists at Seoul National University in Korea using 1992-1998 SCI data. The SCI journals preferred by Korean chemists were the *Bulletin of the Korean Chemical Society (B Kor Chem Soc,* Korea*),*

*Tetrahedron Letters (Tetrahedron Lett*, UK*), Journal of Physical Chemistry A (J Phys Chem A,* USA*), Journal of Chemical Physics (J Chem Phys* , USA*)* and the *Journal of the American Chemical Society (J Am Chem Soc* , USA*)*. This reveals that Korean SCI journals play a role of broadening international publication venues for domestic researchers.

From the perspective of research policy, the internationalization of research is greatly due to the fact that the Korea government started to worry about the relative isolation of the scientific community from the world scientific system. Since the late 1990s, the development strategies and the priorities of research policies in science and technology in Korea can be characterized as internationalization. The year 1998 is generally used to emphasize a decisive break with past eras in Korea. A financial disaster called the 'IMF (International Monetary Fund) crisis' happened because of insufficient currency of US dollars in December 1997. Due to this financial crisis, the Korean government accelerated the development of each sector of society towards more globalization. Prior to the IMF crisis, the academic sector in Korea was more or less adamant about being evaluated by globally acknowledged scientific standard. After the crisis, however, one vigorously adopts an editorial and evaluation system according to the guidelines of international publication practices.

In accordance with globalization trends, the Korean government has undertaken a multi-faceted endeavor including financial assistance to overcome scientific localism. Kim (2005) documented that the R&D expenditure from the private sector has significantly decreased because of the Korean economic downturn. Nevertheless, the Korean government has steadily attempted to enlarge R&D activities. The share of the government sector in total R&D expenditure went up from 23 percent in 1997 to 27 percent in 1998-1999. Furthermore, the Korean government has launched a subsidy program for academic societies as well as university-based research centers since early 2000. The primary purpose of these policies is to assist their international publication activities lest Korea should be scientifically fell behind in the global arena. This is a meaningful effort in relation to the objective to internationalize the research output generated by Korean scientists.

Another example is that the number of Korean journals included in the *Science Citation Index*-expanded version has become considerably higher. In 1993, only three Korean journals were indexed in the database. These were *Journal of the Korean Physical Society (J Korean Phys Soc), Korean Journal of Chemical Engineering (Korean J Chem Eng),* and the *Bulletin of the Korean Chemical Society (B Kor Chem Soc)*. In June 2006, ten Korean journals were indexed in the *Science Citation Index* and twenty-nine journals were newly included in the *Science Citation Index*-Expanded version. Among these journals, nineteen journals have been included since the year 2000 (see Appendix 1). During this process, a new publishing company with a specialization in science and technology journals, Techno-Press, was also established. This company is regarded as a unique entity in terms of its relative independence of traditional societies. Techno-Press is currently publishing five *Science Citation Index*-expanded journals.[1] These trends reflect the slowly changing situation in the national publication system. Furthermore, four academic societies are printing and distributing their journals through international publishing companies. For example, *Molecules & Cells* and *Journal of Biochemistry & Molecular Biology* are being taken care of by Springer, *Current Applied Physics* is being made by Elsevier, and *Geosciences Journal* is being printed by Mary Ann Liebert.

Another issue is whether Korean scholars are citing papers published in Korean SCI journals more actively than international SCI journals. Kim (2004) concluded that this is not the case. The Korean scientific research system including social science is heavily leaning toward the Western practices (e.g., the U.S.A.). However, interesting practices have been developed among Korean researchers and journal editors. Journal publishers have an interest in increasing their citation rates. Therefore, a journal may informally ask prospective contributors to cite at least two articles published in the same journal. In other words, the journal publisher promotes itself by increasing its self-citation rate. Furthermore, some journal publishers provide citing author with cash-coupons. For example, if one publishes an article in a journal included in the Thomson ISI database and cite another journal, one can obtain a coupon as a gratuity from the publishing house of the cited journal. This contribution is dubious, but no longer easily correctable as a within-journal 'self'-citation (Leydesdorff, 2007).

## Methods and Data

### *Social network analysis*

Methodological techniques developed in social network analysis are applied to this research. Network analysis is a set of research procedures for identifying structures in social systems based on the relations among the system's components rather than the attributes of individual cases (Wasserman & Faust, 1994). The method has been previously applied to describe the patterns of scientific communication (Kim, Park, & Thelwall, 2006; Leydesdorff, 2007; Park & Thelwall, 2006). Network analysis is especially useful for identifying individual nodes that are most central (or peripheral) to the citation network. In other words, journals that are the largest information sources and targets of citations can be examined through the use of network analysis.

---

[1] They are *Structural Engineering and Mechanics, Wind & Structures, Steel & Composite Structures, Computers and Concrete, Smart Structures and Systems.*

For example, Freeman's (1979) degree centrality is a basic and primary measure in social network analysis. The indegree centrality of a journal means the connectivity number of journals that are linked to a given journal in terms of citation. On the other hand, outdegree centrality refers to how many citations each journal has created in its reference section. While degree centrality is related to the position of individual nodes, in this case, journals, system indicator centralization (that is, indegree centralization) tells us the extent to which the citation is concentrated to the highly cited journals in the network. The higher this percentage, the more centralized. In other words, there is unequal distribution of citations at the level of the network.

Furthermore, researchers explicitly illustrate the overall citation pattern through a network diagram. The network diagram enables us to show how varied the citation relations of target journals per source journal are. This visualization can simultaneously display the relationship between the citing and cited journals. In these visualizations, circles will represent international journals and squares Korean journals. The size of the journals will be proportional to the number of their being citedness. Lines between them imply the presence of citations and arrow heads indicate the direction of citations. The length of lines does not directly represent the total number of citations but shorter lines tend to represent higher citation counts.

During the visualization process, the average value of citation matrix (also called density in social network analysis) is used as the threshold level. In other words, lines between journals are omitted when the citation count is below the average. In network diagram, this value is frequently employed as the threshold level to make the hidden linkage pattern among nodes more visible (Wasserman & Faust, 1994). For this research, social network analysis is conducted using UCINET for Windows (Borgatti, Everett, & Freeman, 2002). More specifically, a social network visualization technique is applied into the data using the NetDraw available from UCINET for Windows.

*Data*

This paper examines the networked structure of citations of Korean SCI journals. The procedures of data collection are as follows. First, a list of ten Korean SCI journals was made as provided in Table 1. We then obtained their citation records from the *Journal Citation Reports* of the *Science Citation Index* 2004 published by the Thomson ISI. We used the CD-Rom version of these reports.

More specifically, a journal-journal citation matrix is made to acquire a citation frequency from individual Korean SCI journals to all other journals listed in the database. The citations are counted in terms of unique article relations. Citation relations lower than two are aggregated by the ISI under the category “All others.” This operationalization provides us with a global citation environment of the Korean SCI journals.

In the case of the *B Kor Chem Soc,* for example, 198 journals are generated as its reference group in terms of journals containing articles which cite this journal. This procedure enabled us to discover a relational structure of interconnectivity among the journals citing each of Korean SCI journals under investigation. Also we can pinpoint the specific positions of individual Korean journals in their fields by finding out the extent to which journals are citing within their own group compared to how often they are cited inside of their group. We can highlight imbalances in the citation patterns between Korean SCI journals and international SCI journals.

**Results**

*Profile of Korean SCI journals in global citation environment*

Table 2 shows the portfolios of the ten Korean journals which are included in the Science Citation Index, with their global citation environment. The number of journals cited in the second column refers to how many journals in the journal-to-journal citation network of the ISI database in 2004 cited this Korean journal. For example, *B Kor Chem Soc* was cited by 198 journals in the year 2004 while it cited 452 journals. *B Kor Chem Soc* was the most cited journal among the 10 Korean journals. *Mol Cells, J Korean Med Sci,* and *J Korean Phys Soc* followed. *B Kor Chem Soc* was indexed in 1981 by the Thomson ISI and has long been published by a representative organization of Korea in the chemistry field.

In the context of networked research in the information society, the scientific value of journals is largely dependent on the networking capability of the papers published in them. However, there is a big discrepancy between the number of being cited and that of citing for all the Korean SCI journals. All Korean journals were cited by other journals much less than the times they cite the others. This means that their visibility, impact, or reputation in the international scholarly community is weak when the association between journals is measured using aggregated citation relations. In the global citation environment, scientific articles of Korean SCI journals are not regarded as highly trustworthy sources. Being less cited may be caused by the fact that Korean SCI journals hardly have international publishers so that it is relatively difficult for them to get exposed to world scientific community. They have a weak position on the market. The number of non-Koreans publishing in Korean journals is low.

Table 2. Performance of Korean SCI journals in the global citation environment

| No | Journal title | Number of journals citing (Indegree) | Number of journals cited (Outdegree) |
|---|---|---|---|
| 1 | *Bulletin of the Korean Chemical Society* | 198 | 452 |
| 2 | *Journal of the Korean Physical Society* | 129 | 390 |
| 3 | *Molecules and Cells* | 144 | 354 |
| 4 | *ETRI Journal* | 19 | 74 |
| 5 | *Journal of Microbiology and Biotechnology* | 72 | 392 |
| 6 | *Experimental and Molecular Medicine* | 89 | 229 |
| 7 | *Macromolecular Research* | 25 | 209 |
| 8 | *Journal of Communications and Networks* | 13 | 28 |
| 9 | *Journal of Korean Medical Science* | 131 | 451 |
| 10 | *Journal of Ceramic Processing Research* | 11 | 100 |

*Profile of Korean SCI journals in their local citation environments*

In order to examine local citation environment, we collected the population of journals which cited a paper published in the Korean SCI journals. The local impact of a journal can be examined by its share of the total citations in the journal's neighboring citation environment (Leydesdorff, 2007; Leydesdorff & Park, in preparation). Overall, there is a large discrepancy between the number of citing and being cited except for *Etri J*. We do not see a significant number of citations from a reference group of journals that cited Korean SCI journals to the latter.

The *B Kor Chem Soc* is the most influential journal in terms of the number of being cited (1,185 citations). Here we examine *J Microbiol Biotechn*. Among the 10 Korean SCI journals under investigation, this journal has the relatively high visibility in terms of its rank in the cited network. As indicated in Table 3, it occupied the 48th position among the 72 journals. *J Microbiol Biotechn* obtained a total of 265 citations from 72 journals except for self-citations. This tells us that *J Microbiol Biotechn* is fairly acknowledged as a competitive Korean journal in the microbiology field.

Table 3. Performance of Korean SCI journals in local citation environment

| No | Journal title | Nr of publications citing (Indegree) | Rank in terms of being cited (Indegree) | Nr of publications cited (Outdegree) |
|---|---|---|---|---|
| 1 | *Bulletin of the Korean Chemical Society* | 1,185 | 116th among 198 jrns | 5,128 |
| 2 | *Journal of the Korean Physical Society* | 741 | 87th among 129 jrns | 7,334 |
| 3 | *Molecules and Cells* | 537 | 101st among 144 jrns | 1,914 |
| 4 | *ETRI Journal* | 78 | 15th among 19 jrns | 77 |
| 5 | *Journal of Microbiology and Biotechnology* | 265 | 48th among 72 jrns | 1,601 |
| 6 | *Experimental and Molecular Medicine* | 273 | 79th among 89 jrns | 891 |
| 7 | *Macromolecular Research* | 88 | 22nd among 25 jrns | 1,464 |
| 8 | *Journal of Communications and Networks* | 46 | $13^{th}$ among 13 jrns | 134 |
| 9 | *Journal of Korean Medical Science* | 342 | $84^{th}$ among 131 jrns | 752 |
| 10 | *Journal of Ceramic Processing Research* | 23 | $10^{th}$ among 11 jrns | 336 |

**1. *B Kor Chem Soc***

a. Complete network of *B Kor Chem Soc* in the cited dimension

Using *B Kor Chem Soc* as the entrance journal, we made a citation network as seen in Figure 2. The results reveal that *J Am Chem Soc* (145,333 out of 1,215,303 citations) is the most central journal in this network. Next, *J Chem Phys* (57,121 citations), *J Org Chem* (52,128 citations), and *Angew Chem Int Edit* (50,456 citations) are the next most central journals (Bornmann, Leydesdorff, & Marx, forthcoming). These journals among other journals are located in the center of the map. Although this network is made based on *B Kor Chem Soc,* academic literature in international journals has an impact on whipping the structure of citation network in a shape. Most peripheral in the network are *Curr Org Synth* (2 citations), *B Chem Soc Ethiopia* (4 citations), and *Concept Magn Reson A* (6 citations). *Cent Eur J Chem* was isolated from the network without receiving any citation from other journals. The least cited journals in the group are scattered around in the map.

b. Ego-network of *B Kor Chem Soc* in the cited dimension

*B Kor Chem Soc* is selected as a focal journal. The *J Org Chem* (63), *Tetrahedron Lett* (45), *Tetrahedron* (39), *Syn Lett* (30), and *J Am Chem Soc* (21) frequently cited papers published in *B Kor Chem Soc*. However, the number of citations from *B Kor Chem Soc* to international journals is much higher than the reverse citation relation: *J Am Chem Soc* (632), *J Org Chem* (432), *Tetrahedron Lett* (353), and *J Chem Phys* (182). This shows that leading journals in the international scientific community provide the largest sources of scientific knowledge for Korean academics.

How is the citation situation among Korean SCI journals in the network? *B Kor Chem Soc* received only 41 citations (out of 1,185 citations) from papers published in other Korean journals. These are *Macromol Res* (14), *J Ind Eng Chem* (12), *Arch Pharm Res* (4), *J Microbiol*

*Biotechn*(4), *Korean J Chem Eng* (4), *Polym-Korea* (3). Similar to the number of being cited by domestic journals, *B Kor Chem Soc* cited their publications 35 times: *J Ind Eng Chem* (14), *Arch Pharm Res* (8), *Korean J Chem Eng* (6), *Macromol Res* (5), *J Microbiol Biotechn* (2). *Polym-Korea* was not cited at all by *B Kor Chem Soc*. *B Kor Chem Soc* is not firmly integrated with domestic journals in the Thomson ISI database. The former is positioned far from the latter in Figure 2. This means that *B Kor Chem Soc* does not play a role of a hub among marginal Korean journals in this field. In other words, Korean chemists are eager to acquire research output generated by their international counterparts and give them credit. Interestingly, *J Microbiol Biotechn* and *Arch Pharm Res*, *Polym-Korea* and *Macromol Res*, *J Ind Eng Chem* and *Korean J Chem Eng* are arranged on the rim side of Figure 2 by making a pair. This indicates that these journals are citing international journals in a similar pattern.

Figure 2. Network diagram among the 198 journals citing *B Kor Chem Soc*

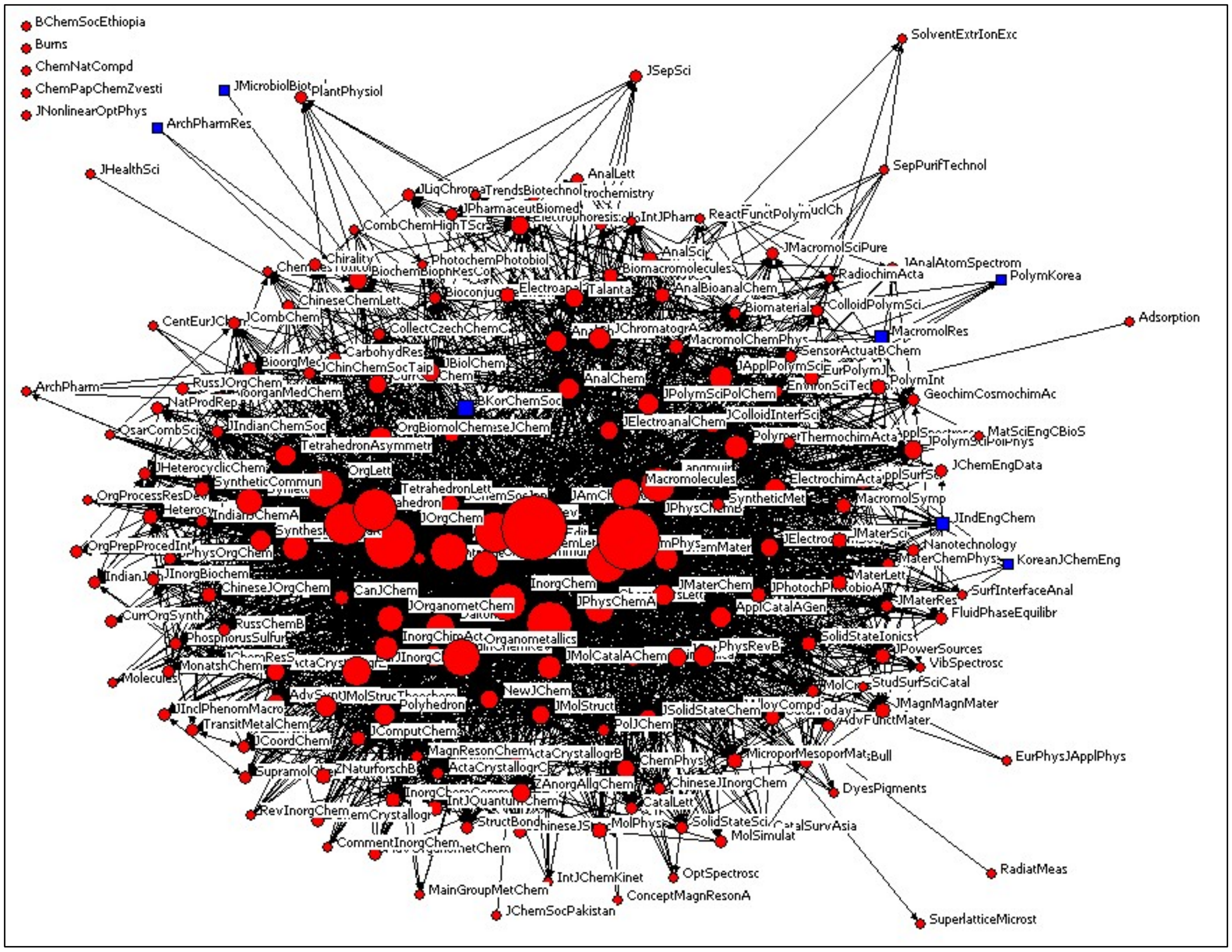

**2. *J Korean Phys Soc***

a. Complete network of *J Korean Phys Soc* in the cited dimension

In the network diagram of the citation environment of *J Korean Phys Soc*, the most cited journal *Phys Rev Lett* (133,631) is very visible in the middle of Figure 3. The next highly cited journals *Phys Rev B* (88,362), *Appl Phys Lett* (58,110), *J Appl Phys* (44,205), *Phys Lett B* (36,533), and *Phys Rev D* (32,647) are also fairly noticeable in the map. The peripheral journals in the network are *Etri J* (44), *Iau Symp* (15), *Cr Chim* (14), and *Asian J Spectrosc* (3).

b. Ego-network of *J Korean Phys Soc* in the cited dimension

In the case of citations of *J Korean Phys Soc*, *Jpn J Appl Phys* (49) cited it the most. *Ferroelectrics* (34) and *J Appl Phys* (32) also cited *J Korean Phys Soc* more than 30 times. Among Korean journals in the group, *Curr Appl Phys* cited *J Korean Phys Soc* 16 times. Both *Curr Appl Phys* and cited *J Korean Phys Soc* are published by the same organization, Korean Physical Society. *Etri J* and *B Kor Chem Soc* were 4 and 2 times respectively.

Figure 3. Network diagram among the 129 journals citing *J Korean Phys Soc*

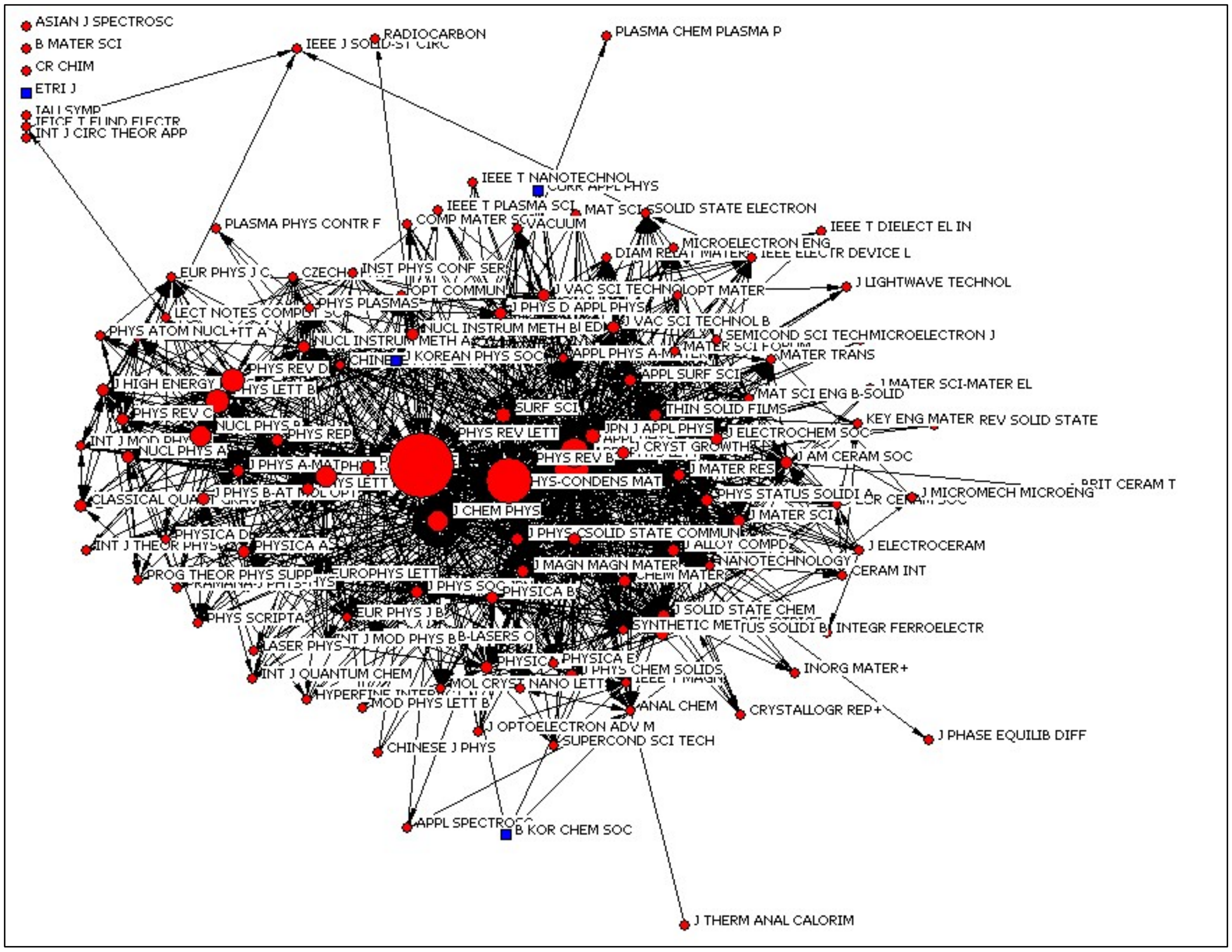

**3. *Mol Cells***

a. Complete network of *Mol Cells* in the cited dimension

We made a network-diagram mapping of citation connections among the 144 journals citing the seed journal, in this case, *Mol Cells*. As seen in Figure 4, a network diagram provides useful indicators that there are two major journals, *J Biol Chem* and *P Natl Acad Sci USA*, in this data. *J Biol Chem* and *P Natl Acad Sci USA* received 125,062 and 109,461 citations respectively from other journals in the group. Interestingly, both journals were also the most central in the network of *J Microbiol Biotechn* as described in the above. The next most cited journals *Embo J* (37,836), *Bio Chemistry-US* (32,874), and *Mol Cell Biol* (30,886) are neatly interspersed making a circle centered on the major journals. *Korean J Genetic* (9), *Cytom Part A* (8), *Asian Austral J Anim*(8), *J Clin Biochem Nutr* (8) were cited less than 10 times. *Chem Biodivers* had none of citations in the group. The map clearly partitions relatively peripheral groups from central ones in international citation landscape of *Mol Cells*.

b. Ego-network of *Mol Cells* in the cited dimension

Let us focus on the disciplinary citation environment using *Mol Cells* as the focal journal. A total of 537 papers in 143 journals included the publications of *Mol Cells* in their references. Major journal *J Biol Chem* cited *Mol Cells* the most (44 times) and *Biochem Bioph Res Co* was the second (26 times). *Nucleic Acids Res* (19), *Free Radical Bio Med* (13), *Korean J Genetic* (13), *Plant Physiol* (12), *J Microbiol Biotechn* (11), and *Plant J* (10) cited *Mol Cells* more than 10 times. The number of citations of the eighty journals in this group is 2. In other words, the research studies *Mol Cells* had published were listed 2 times as the reference literature of the 80 journals. This can be a sign of *Mol Cells* being closer to a global journal.

However, *Mol Cells* was cited only 51 times by eight Korean journals in the group: *Korean J Genetic* (13), *J Microbiol Biotechn* (11), *Exp Mol Med* (8), *J Microbiol* (6), *J Biochem Mol Biol* (5), *Arch Pharm Res* (4), *Asian Austral J Anim* (2), and *Yonsei Med J* (2). This reveals that Korean scholars like to make a collaborative weaving with the scientists publishing their research in international journals rather than domestic counter partners. Additionally, it is interesting to see that the eight Korean journals are divided in three clusters in the Figure 4. This could have happened owing to their similar citing practices.

Figure 4. Network diagram among the 144 journals citing *Mol Cells*

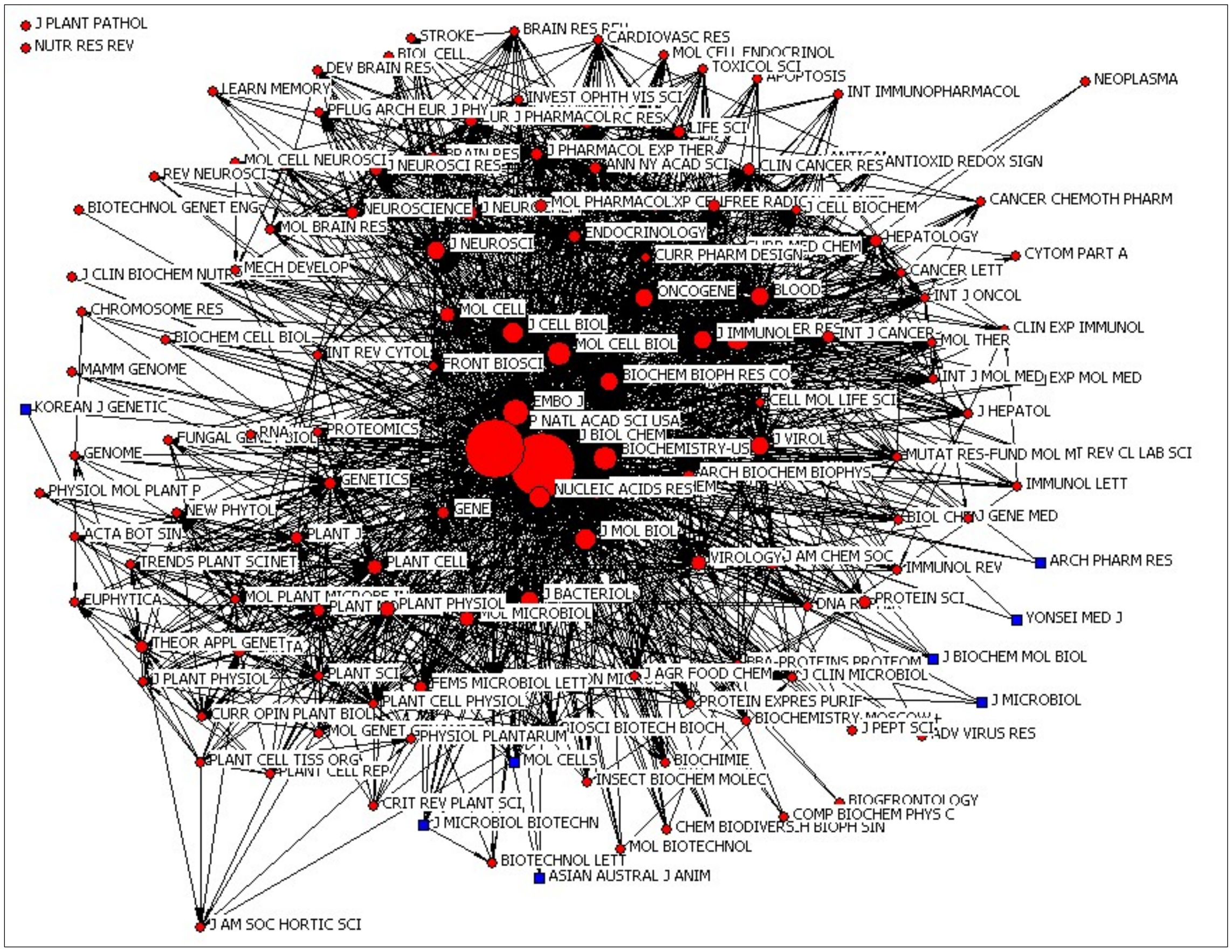

## 4. *Etri J*

a. Complete network of *Etri J* in the cited dimension

In the complete network of *Etri J, Opt Lett* received 2,472 citations from the 19 journals. *IEEE Photonic Tech L* (1,479 citations) and *Jpn J Appl Phys* (1,094 citations) came after *Opt Lett*. Contrary to these central journals, the most peripheral journal *Phys Chem Chem Phys* had only two citations.

b. Ego-network of *Etri J* in the cited dimension

Among the ten Korean SCI journals, *Etri J* is only journal that the number of being cited (78 times) is bigger than that of citing (77 times) in local citation environment. Given that articles published in *Etri J* receive several citations from papers in its reference journals, the journal seems to be perceived as highly reputable compared to the other nine Korean SCI journals.

However, on the close examination of citation counts show that *Etri J* was mostly cited by *Lect Notes Comput Sc* (also called *LNCS*) 28 times (out of 78 citations). *LNCS* is little bit different from regular ISI-ranked journals. Internationally prestigious conference proceedings in science and technology are often published in *LNCS*. However, because these are conference proceedings, it is considered as relatively easy to publish papers in *LNCS*. For this reason, some major universities in Korea do not acknowledge the *LNCS* publications as ISI-rated research performance of their faculty members. Thus, it might not be fair to recognize a number of citations from papers contained in the *LNCS* series as genuine count for a comparison purpose. In addition, there is one Korean journal (*J Korean Phys Soc*) in this network. *Etri J* was cited 4 times by *J Korean Phys Soc*.

**5. *J Microbiol Biotechn***

a. Complete network of *J Microbiol Biotechn* in the cited dimension

First, let's examine a whole citation network with a seed journal of *J Microbiol Biotechn*. The publications of *J Biol Chem* (35,378) and *P Natl Acad Sci USA* (34,863) are the most referenced research in the group. Figure 5 illustrates that these two journals are the hubs of the network in the subject area of *J Microbiol Biotechn*. Both journals are being surrounded by the next cited journals, *Biochemistry-US* (16,121) and *J Bacteriol* (14,421). On the other hand, Korean SCI journals are rather away and isolated from the nucleus of the network. This implies that the majority of Korean scientific knowledge has been input through quality papers published in international journals. The least cited journals are *Compost Sci Util* (29), *J Microbiol* (27), *Korean J Chem Eng* (27), *Key Eng Mater* (6). These peripheral journals are dispersed around in network diagram.

b. Ego-network of *J Microbiol Biotechn* in the cited dimension

Then we move to assess the citation performance of *J Microbiol Biotechn* in the international environment. This journal is between core group and peripheral journals. *J Microbiol Biotechn* was referred to more than 10 times by the authors of non-Korean journals: *Enzyme Microb Tech* (16), *Appl Environ Microb* (10), and *Biotechnol Lett* (10). This citation exchange makes these journals clustered together on the left side of central part. Further, the distribution of citations of *J Microbiol Biotechn* is fairly even among the sample journals as summarized in Table 4. About half of the journals (49.3%) cited *J Microbio Biotechn* more than 2 times. Although *J Microbiol Biotechn* has not yet had a substantial impact on international scholars in the relevant field, the journal's visibility in the global scientific community is to some extent gained by a number of small contributions.

*J Microbiol Biotechn* received only 20 citations from the 4 domestic counterparts in the group: *J Microbiol* (12), *J Biochem Mol Biol* (4), *B Kor Chem Soc* (2), and *Korean J Chem Eng* (2). In particular, the sixty percent of domestic citations came from *J Microbiol* published by a peer organization (Korean Society for Microbiology) of the publisher of *J Microbiol Biotechn* (Korean Society for Microbiology and Biotechnology). In the meantime, *J Microbiol Biotechn* cited *J Microbiol, J Biochem Mol Biol*, and *B Kor Chem Soc* in 13, 8, and 4 times respectively. The citation *to Korean J Chem Eng* was not existent at all.

Table 4. Distribution of *J Microbiol Biotechn* citations

| Nr of citations | Nr of journals | Percent |
|---|---|---|
| 2 | 35 | 49.3 |
| 3 | 11 | 15.5 |
| 4 | 10 | 14.1 |
| 5 | 4 | 5.6 |
| 6 | 2 | 2.8 |
| 8 | 3 | 4.2 |
| 9 | 2 | 2.8 |
| 10 | 2 | 2.8 |
| 12 | 1 | 1.4 |
| 16 | 1 | 1.4 |
| Total | 71 | 100.0 |

Figure 5. Network diagram among the 72 journals citing *J Microbiol Biotechn*

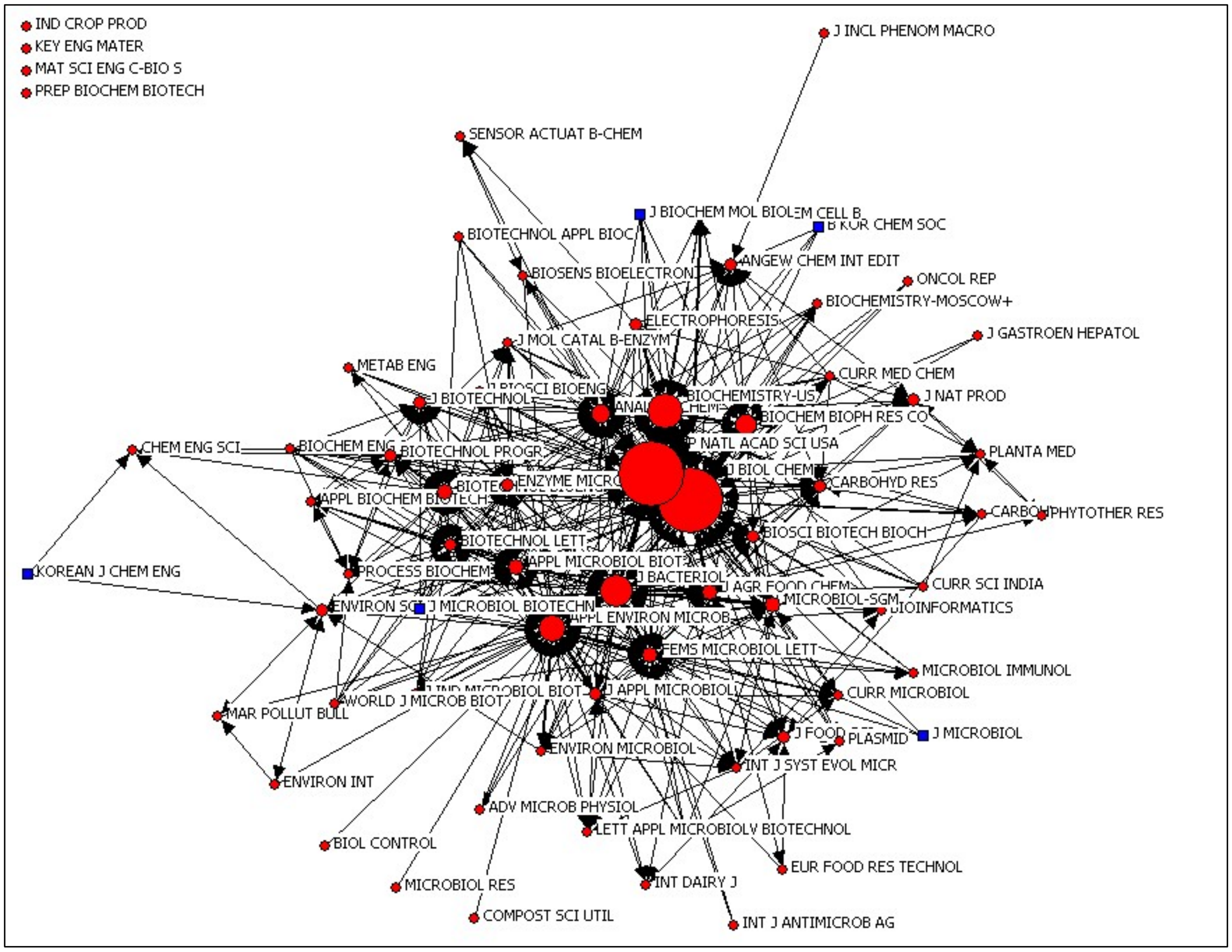

## 6. ***Exp Mol Med***

a. Complete network of *Exp Mol Med* in the cited dimension

Using *Exp Mol Med* as a starting point, a network composed of 89 journals was drawn. In this network, *J Biol Chem* is extremely cited journal by others (101,317 citations). The next frequently cited journals are *Cancer Res* (31,735) and *Mol Cell Biol* (25,144). On other hand, *Hepatol Res* (50 citations) and *B Exp Biol Med+* (26 citations) were marginal journals in the cited network.

b. Ego-network of *Exp Mol Med* in the cited dimension

In the network of *Exp Mol Med* in the cited dimension, *Exp Mol Med* was cited 33 times by *J Biol Chem* that is a hub journal in local citation environment. *Biochem Bioph Res Co* cited *Exp Mol Med* 15 times. On the other hand, only Korean journal in this group *Mol Cells* cited *Exp Mol Med* 2 times.

**7. *Macromol Res***

a. Complete network of *Macromol Res* in the cited dimension

We have drawn a citation network using *Macromol Res* as the seed journal. Figure 6 shows the whole configuration of interconnectivity among journals in a network composed of the journals citing *Macromol Res*. The thickness of lines indicates the frequency of citations that is regarded as the strength of knowledge transfer relations. The journals with high number of citations are assorted in the middle of network map and frequently citing and cited journals are clear in Figure 6. The relatively lesser cited journals are scattered on side.

b. Ego-network of *Macromol Res* in the cited dimension

According to data analysis, *Macromolecules* (19,946) was cited most by the other journals within this group. *Polymer* (10,446) occupied the second position. The next group of most cited journals contains the *J Appl Polym Sci* (5,679), *J Polym Sci Pol Chem* (4,766), and *Langmuir* (4,107). Similar to *B Kor Chem Soc*, *Macromol Res* cited the publications of these international journals (*Macromolecules*, 381; *Polymer*, 211; *J Appl Polym Sci*, 206; *J Polym Sci Pol Chem*, 126) much more than the research reports contained in domestic journals (*Polym-Korea*, 71; *B Kor Chem Soc*, 14; *J Ind Eng Chem*, 5). In the Figure 6, Korean journals are heavily connected to international journals. This means that knowledge printed at international journals with high impact are the primary information sources and communication channels for scientific awakening among Korean academics.

Let us explore the role of *Macromol Res* within the Korean group. Among the 25 journals citing *Macromol Res*, there are three domestic journals. The total times of cited are 20: *Polym-Korea* (12), *B Kor Chem Soc* (5), and *J Ind Eng Chem* (3). This accounts for a 22.7 percent of the citations given to *Macromol Res* (20 out of 88 citations). Interestingly, *Polymer-Korea* has a close relation with *Macromol Res* compared to other domestic journals. This is partly due to the fact that two journals are published by the same institution, that is, the Polymer Society of Korea. One can conjecture that this may provide an institutional incentive. We suspect that there might be some self-citations at individual author's level and organizational efforts could be involved for the development of both journals. Given that the authors in *Macromol Res* are citing *Polym-Korea* (citing 71 times vs cited 12 times), *Polym-Korea* has a potential to grow as a more prestigious journal than now and may be indexed in the SCI database, beyond the latter's expanded version.

Figure 6. Network diagram among the 25 journals citing Macromolecular Research

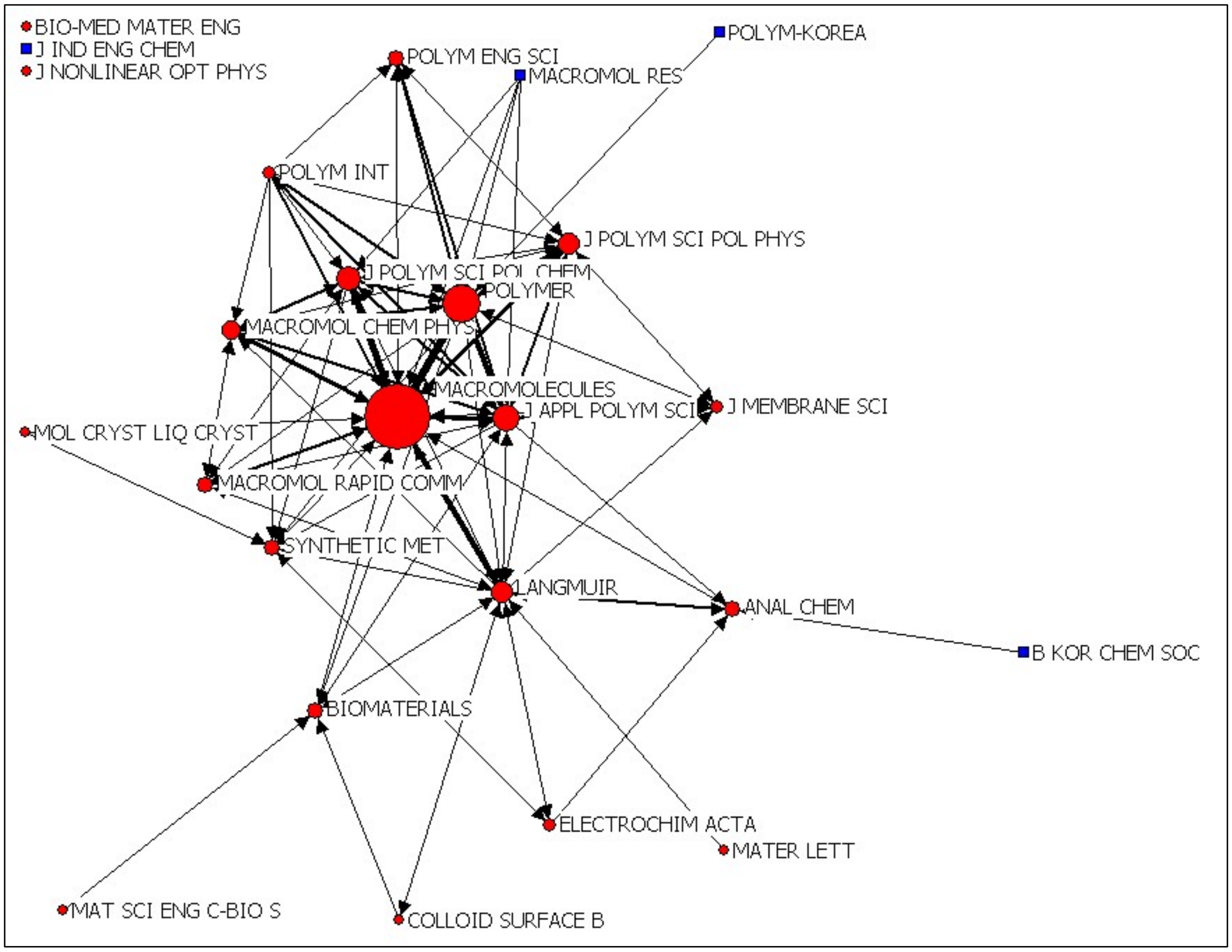

**8. *J Commun Netw - S Kor***

a. Complete network of *J Commun Netw-S Kor* in the cited dimension

The citation network analysis of *J Commun Netw-S Kor* reveals a particular map of relationships where *IEEE J Sel Area Comm* owns the largest percentage of total citations (2,264 out of 7,326, 30.90%) followed by *IEEE T Commun* (2,252, 30.74%). They seem to offer basic information for any scientists to catch up with the cutting-edge trends in this field. Their publications work as entry point into a greater scientific discourse. The seed journal *J Commun Netw-S Kor* is also citing the two journals the most. *J Commun Netw-S Kor*'s citation to top two journals may be related to specific purpose and goals. *J Commun Netw-S Kor* is the least cited journal in the network. The authors of *J Commun Netw-S Kor* need to enhance their perceived credibility by connecting and networking with papers published in major journals. Top journals offer a number of highly-selective papers that can be a source by authors for marginal journals.

b. Ego-network of *J Commun Netw-S Kor* in the cited dimension

According to the analysis of ego-network of *J Commun Netw-S Kor* in the cited dimension, *J Commun Netw- S Korea* was cited by *LNCS* the most (9 out of 46). Next, *IEEE Signal Proc Mag* and *IEEE T Wirel Commun* cited *J Commun Netw- S Korea* 7 times respectively.

**9. *J Korean Med Sci***

a. Complete network of *J Korean Med Sci* in the cited dimension

In aggregated journal-journal citation networks based on the seed journal of *J Korean Med Sci,* we are easily able to discern the citation position of a scientific journal, for example, in terms of core group and peripheries. Journals at core are *P Natl Acad Sci USA* (38,294 citations), *Cancer Res* (26,025 citations), and *J Immunol* (14,827 citations). The least cited journals, that is, journals at periphery include *J Pharmacol Sci* (10 citations), *Saudi Med J* (4 citations), and *Acta Med Aust* (4 citations). *Exp Oncol* gained none of citations.

b. Ego-network of *J Korean Med Sci* in the cited dimension

Using *J Korean Med Sci* as an entrance journal, 131 journals are generated as a reference group. However, none of Korean journals are existent in this network. *J Korean Med Sci* has 342 citations from international journals in this field. A total of 131 journals cited at least one publication in *J Korean Med Sci* but there is not a heavily citing journal. The most citing journals are six journals including *Am J Clin Pathol* that cited *J Korean Med Sci* six times.

**10. *J Ceram Process Res***

a. Complete network of *J Ceram Process Res* in the cited dimension

In a citation network employing *J Ceram Process Res* as a beginning journal, *J Am Ceram Tech* is by far the strongest journal in terms of citation performance. The journal received 5,123 citations from some publications in the remaining journals in the group. The next highly cited journals in the network are *J Mater Sci* (1,754 citations), *J Eur Ceram Soc* (1,322 citations), and *J Mater Res* (1,166). However, the difference between the most quoted journal *J Am Ceram Tech* and the next referred journals is huge. Such a big gap in the number of citations makes this network very centralized around a hub journal. The indegree (that is cited) centralization of this network is 30.52 percent, which is the highest among ten networks. The seed journal *J Ceram Process Res* is the least cited journal in this network. *J Physiol-London* is isolated without any citation coming to itself.

b. Ego-network of *J Ceram Process Res* in the cited dimension

None of Korean journals appeared in the ego-network of *J Ceram Process Res*. While *J Ceram Process Res* cited eleven journals 336 times, they cited back *J Ceram Process Res* only 23 times. *J Ceram Process Res* was cited by *Ceram Int*, *J Eur Ceram Soc*, and *Key Eng Mater* three times respectively.

## Conclusions

The detailed analysis of the citation patterns of these ten Korean journals has taught us that these journals function each of them in specific niches of the international literature. Thus, they may provide Korean scholars with access to these literatures by facilitating the respective entry barriers. Furthermore, they function for organizing the "national subfields" by the professional association that is organizing these researchers at the national level. Among them, however, there are no citation relations based on their Korean background. Thus, there seems nothing specific "Korean" about these ten journals.

The risk of such journals is that they function as second-rate international journals. Unlike Chinese journals (Zhou & Leydesdorff, 2007), Korean journals with international ambition no longer use the national language. In other countries (Japan, France, Germany, Russia) one still finds remainders of national research systems using the respective national languages, but more so in the humanities and the social science than in the natural and life sciences. Given the international aspirations of the Korean journals, they may be well-advised to integrate more in the international publication structure, for example, via mergers or acquisitions. This would also ease the problem of the financing of these journals by professional associations.

A country like the Netherlands, for example, which hosts a number of international publishing houses has virtually no journals included in the ISI-database which are specifically entitled as Dutch. Israel is another example of a small nation with a strong scientific profile and with no specific journals at the international level. The functional division of labor between newsletters of societies (in the national language) and international journals (in English) is dampened in Korea by this layer of national journals in English and competing for enlistment by the ISI. Given their asymmetric citation profiles, it is questionable whether Korean scientists should be advised to submit their work to these journals rather than to truly international ones. Given their status, however, of currently being enlisted in the ISI, it would be functional to keep their names as flagships of the Korean national science system.

As a policy implication, the Korean government should encourage more journals to follow the path of these ten journals and perhaps facilitate the emergence of new journals within the Korean context with international ambition. These journals have lower entrance barriers for Korean scholars, provide an additional circuit of communication, while being keenly aware of their own marginality in terms of the international system.

**Discussions**

Thanks to the public and private sector's commitment to broaden international publications, South Korea has particularly shown a rapid increase in its academic performance. This portfolio also provides us with a new perspective on the evaluation of the spectacular increase of Korean publications indexed in the *Science Citation Index* during the last ten years. We have already seen linear growth in the case of other OECD countries, for example, Scandinavian countries and the Netherlands during the 1980s and Mediterranean countries during the 1990s. However, these seem to have been transition periods during which these countries increasingly began to be integrated into the international system.

In spite of the rapid increase of the number of Korean papers published in *Science Citation Index*, Korean SCI journals occupied a marginal position in their local citation environment. Korean academic societies and policy agencies have been quite successful in making several Korean journals included in the internationally-acknowledge scholarly database. However, it is questionable that these journals have been lived up to its promises in terms of citation performance. In a citation network using Korean SCJ journals as seed journals, Korean titles have low international visibility and authors in international journals hardly quote papers published in these Korean journals. It seems from our research that the Korean SCI journals function more like the German and French journals, that is, at the margin of the international journal system.

Korean journals distinguish themselves from German and French publications in that English is their official language. The use of English also makes these journals different from Chinese journals. According to Leydesdorff and Jin (2005), Chinese journals provide a kind of "Mode 2" structure, for example, in integrating knowledge about political and social priorities like health, agriculture, etc. In citation network diagram, there also exists a geographical grouping among Chinese journals. In contrast to Chinese patterns, Korean SCI journals do not make an independent cluster because of the lack of inter-citations among themselves in the diagrams. Nevertheless, the rise in publications shows some "Chinese" patterns.

Arguably, Korean academic societies have run national journals for political reasons because of the supply of domestic funding for themselves. In Korea, there is also a "Chinese" (Asian) mechanism at work: Every professional society wishes to have its own ISI-rated international journal in order to establish its national elite-circle. Publishing an ISI journal is often indicative of top societies in Korea. This is in contrast to Chinese journals that do not manage to break through the circle of Chinese publications even if they are international and included in the SCI.

Furthermore, the results reveal that Korean SCI journals are not being selected as a serious communication channel among national researchers. Korean SCI journals are being neglected by national scientists as a reference journal. These journals function more like publication places, neither research channels nor information sources. In other words, a few prominent international journals are playing a dominant role of producing knowledge that Korean scientists actively seek to absorb. Thus, the Korean publication system could be better depicted as an institutional system, not as a communication system. This is different from Chinese journals that play a role of a broker between international publications and national scientists in scientific communication.

However, not all the papers available at high impact journals are being taken up equally, which implies that some research output of Korean journals at the periphery could certainly be a rich, diverse, engaging, and stimulating resource in other circumstances. The suggested advices for Korean journal editors and publishers include the followings. Korean journals (a) need to carry more socio-cultural values (e.g., a themed issue for specific geographical or disciplinary orientation), (b) attempt to contain problem-solving knowledge in local research context (e.g., an informational section to diffuse knowledge of global research practice and to increase awareness of trouble-shooting strategies for new-comers in international scholarly community), (c) should be equipped with efficient distribution channel to survive in highly-competitive market (e.g., online archive with a collection of related Korean research papers).

Despite the speedy increase in Korea's publishing SCI papers, the average value of citation per article has not been substantially going up except for some special areas. In terms of policy implications, this situation suggests that the Korean government is recommended to provide appropriate opportunities within a range of graduate curriculum areas to teach how to emphasize and highlight research result in English. Furthermore, the Korean research policy institute could take the initiative to launch an international alliance network specializing in scientific communication and scholarly publishing towards scientists.

**Appendix**. A profile of the 29 Korean SCI-Expanded journals

| No. | JOURNAL TITLE | Since when included in the SCI-Expanded |
|---|---|---|
| 1 | Archives of Pharmacal Research | 1995 |
| 2 | Asian Australasian Journal of Animal Sciences | 1997 |
| 3 | Journal of Biochemistry and Molecular Biology | 1995 |
| 4 | Journal of Industrial and Engineering Chemistry | 1998 |
| 5 | Journal of Microbiology | 1995 |
| 6 | Korean Journal of Chemical Engineering | 1988 |
| 7 | Korean Journal of Genetics | 1995 |
| 8 | Korean Journal of Radiology | 2001 |
| 9 | KSME International Journal[2] | 1995 |
| 10 | Metals and Materials International | 1997 |
| 11 | Polymer-Korea | 1995 |
| 12 | Structural Engineering and Mechanics | 1995 |
| 13 | Yonsei Medical Journal | 1998 |
| 14 | Wind & Structures | 2000 |
| 15 | Current Applied Physics | 2002 |
| 16 | Korea-Australia Rheology Journal | 2002 |
| 17 | Journal of the Korean Mathematical Society | 2003 |
| 18 | STEEL & COMPOSITE STRUCTURES | 2003 |
| 19 | Fibers and Polymers | 2003 |
| 20 | International Journal of Automotive Technology | 2003 |
| 21 | Food Science and Biotechnology | 2003 |
| 22 | Journal of Plant Biology | 2003 |
| 23 | Biotechnology and Bioprocess Engineering | 2003 |
| 24 | Geosciences Journal | 2003 |
| 25 | Journal of Medicinal Food | 2004 |
| 26 | International Journal of Control Automation and Systems | 2004 |
| 27 | Computers and Concrete | 2005 |
| 28 | Smart Structures and Systems | 2005 |
| 29 | Journal of Veterinary Science | 2006 |

[2] KSME International Journal was changed to Journal of Mechanical Science and Technology in January 2005.

* Note: Two of the above journals (Korean Journal of Genetics and Polymer-Korea) allow articles to be submitted in the Korean language.